\documentclass[aps,prd,twocolumn,floatfix,preprintnumbers,nofootinbib,superscriptaddress]{revtex4}
\usepackage{mathtools}
\usepackage{graphicx}
\usepackage{color}
\usepackage{natbib}
\usepackage{multirow}
\usepackage{amsmath}
\usepackage[amssymb]{SIunits}

\newcommand{\dd}{\mathrm{d}}

\begin{document}

\title{CMB bounds on primordial black holes including dark matter halo accretion}
\author{Pasquale D. Serpico}\email{serpico@lapth.cnrs.fr}
\affiliation{Univ. Grenoble Alpes, USMB, CNRS, LAPTh, F-74940 Annecy, France}
\author{Vivian Poulin}
\affiliation{LUPM,  CNRS  \&  Universit\'e  de  Montpellier, F-34095 Montpellier, France}
\author{Derek Inman}
\affiliation{Center for Cosmology and Particle Physics, Department of Physics, New York University, 726 Broadway, New York, NY, 10003, USA}
\author{Kazunori Kohri}
\affiliation{Theory Center, IPNS, KEK, Tsukuba 305-0801, Ibaraki, Japan}
\affiliation{The Graduate University of Advanced Studies (Sokendai), Tsukuba 305-0801, Ibaraki,Japan}
\affiliation{Rudolf Peierls Centre for Theoretical Physics, University of Oxford, Department of Physics, Oxford, OX1 3PU, UK}
\affiliation{Kavli IPMU (WPI), UTIAS, The University of Tokyo, Kashiwa, Chiba 277-8583, Japan}

\date{\today}

\preprint{LAPTH-005/20, KEK-Cosmo-248, KEK-TH-2198, IPMU20-0021}

\begin{abstract} 
Even if massive ($10\,M_\odot \lesssim M \lesssim 10^4 M_\odot$) primordial black holes (PBHs) can only account for a small fraction of the dark matter (DM) in the universe, they may still be responsible for a sizable fraction of the coalescence events measured by LIGO/Virgo, and/or act as progenitors of the supermassive black holes (SMBHs) observed already at high redshift ($z\gtrsim 6$). In presence of a dominant, non-PBH DM component, the bounds set by CMB via an altered ionization history are modified. 
We revisit the cosmological accretion of a DM halo around PBHs via toy models and dedicated numerical simulations, deriving updated CMB bounds which  also take into account the last Planck data release. We prove that these constraints dominate over other constraints available in the literature at masses $M\gtrsim 20-50\,M_\odot$ (depending on uncertainty in accretion physics), reaching the level $f_{\rm PBH}<3\times 10^{-9}$ around $M\sim 10^{4}\,M_\odot$. These tight bounds are nonetheless consistent with the hypothesis of a primordial origin of the SMBH massive seeds.
\end{abstract}

\maketitle

\section{Introduction}
Despite several decades of  direct, indirect or collider searches, we are still ignorant of the nature of mark matter (DM) of the universe. 
It is even unclear if it is made of a single species or if, just like its baryonic matter counterpart, it is constituted by different components. Based on current constraints, primordial black holes (PBH) formed in the early universe remain a viable DM candidate only in the mass window $10^{-16}\,M_\odot \lesssim M \lesssim 10^{-11}\,M_\odot$ (see e.g.~\cite{Ballesteros:2020qam,Kusenko:2020pcg,Carr:2020gox}). However, PBH of much higher masses, even if not constituting the bulk of DM, can still have other interesting cosmological and astrophysical consequences. For instance, PBH with masses of tens of $M_\odot$ could be responsible for some if not most of the heavy BH mergers discovered by LIGO/Virgo, even if they only contribute a fraction $f_{\rm PBH}\sim {\rm few}\times 10^{-3}$ to the DM~\cite{Sasaki:2016jop,Ali-Haimoud:2017rtz,Kavanagh:2018ggo}. 

One wonderful probe for such putative pristine objects is the cosmic microwave background (CMB). PBH would start accreting gas soon after matter-radiation equality ($z\simeq 3500$); the gas heats up to the point that ionizing radiation is emitted, which alters the opacity of the gas in the long period between recombination and reionization. CMB temperature and polarization fluctuations are extremely sensitive to this phenomenon and can be used to constrain it. This argument has been used by several authors in the past, from the pioneering study~\cite{Ricotti:2007au} to the amended and more recent calculations in Ref.~\cite{Ali-Haimoud:2016mbv,Horowitz:2016lib,Chen:2016pud,Poulin:2017bwe,Bernal:2017vvn}. 

In this article, we extend our previous calculation~\cite{Poulin:2017bwe} to account for the accretion of the dominant, non-PBH DM particles onto PBH,  which enhances 
 the baryonic accretion and eventually the PBH luminosity constrained by the CMB,  simply as a result of the increased gravitational potential felt by the baryons. Besides providing a more realistic assessment of the CMB bounds on stellar-mass PBH, we are also motivated by the possibility that intermediate mass PBH ($10^2\,M_\odot \lesssim M \lesssim 10^{4} M_\odot$) may provide the seeds leading to the super-massive black holes (SMBHs) observed at redshift $z\gtrsim 6$ (with the current record-holder of $10^{8.9}\,M_\odot$ at $z=7.54$~\cite{Banados:2017unc}) whose origin constitutes a long-standing mystery, see e.g.~\cite{Inayoshi:2019fun}. To the best of our knowledge, nobody has assessed the viability of this hypothesis in the light of current CMB anisotropy bounds.

This article is structured as follows. 
In Sec.~\ref{CMBprim}, we start by reviewing the formalism and hypotheses entering the bounds on PBHs set by the CMB. 
Since the main systematic uncertainty in these bounds consists in the treatment of accretion physics, we derive the bounds in a couple of physically motivated benchmarks, which should bracket the uncertainties. 
In Sec.~\ref{loophole}, we discuss the role played by DM halos accreting around PBH, treating them both via semi-analytical toy models and dedicated numerical simulations.
Our CMB bounds are reported in Sec.~\ref{constraints}, providing an upgrade to the the bounds presented in Ref.~\cite{Poulin:2017bwe} in the light of the last Planck data release and the updated treatment of the energy release via the new \texttt{ExoCLASS} package~\cite{Stocker:2018avm}. Also, we present an extension of the bounds up to masses of  $M\simeq {\rm few}\times 10^4\, M_\odot $, beyond which some working hypotheses break down.
Sec.~\ref{SMBH} briefly reviews the puzzle concerning the origin of SMBH, 
and discusses the implications of our CMB limits. We can anticipate that, even under Eddington accretion conditions for the PBH surrounded by DM halos, CMB angular power spectra are not capable of testing the hypothesis that the SMBH detected already at high-redshift are (seeded by) PBH. Hence this remains a viable conjecture with  interesting consequences for the cosmology of the dark ages on which we briefly comment upon in Sec.~\ref{concl}, where we also report our conclusions. 

\section{The luminosity of accreting PBH}\label{CMBprim}
In the following, we assume that the ionization of gas in the dark ages due to accretion onto PBHs and probed by the CMB can be considered homogeneous. A justification is provided in Appendix~\ref{ionsizecheck}.
The key input needed to compute the CMB bound is the total energy injection rate per unit volume:
\begin{equation}\label{eq:dQdt}
\frac{\dd^2 E}{\dd V \dd t}=L_{\rm acc}n_{\rm PBH}=L_{\rm acc} f_{\rm PBH}\frac{\rho_{\rm DM}}{M}\,,
\end{equation}
where $L_{\rm acc}(M,t)$ is the bolometric (in general, time dependent) accretion luminosity onto a PBH of mass $M$, and is the main unknown. A milder uncertainty (within a factor $\lesssim 2$) comes from the spectral distribution of the emitted radiation, which in turns determines the energy {\it deposited} in the medium---what actually matters---and for which we make use of the transfer functions\footnote{More recent tools have been developed for the computation of the energy deposited in the cosmological plasma during the cosmic dark ages \cite{Liu:2019bbm}, but it has been shown that the effect of energy injection onto the CMB bounds is accurately described by the tabulated transfer functions.} from Ref.~\cite{Slatyer09,Slatyer15-2} as implemented in  the \texttt{ExoCLASS} package~\cite{Stocker:2018avm}. As in Ref.~\cite{Poulin:2017bwe}, we assume that the energy-differential spectrum of $L_{\rm acc}$, $L_\omega$, is dominated by Bremsstrahlung emission (see~\cite{1973ApJ...180..531S,1973ApJ...185...69S,Ali-Haimoud:2016mbv}), with a mildly decreasing frequency dependence over several decades, and a cutoff given by the temperature of the medium near the Schwarzschild radius $T_s$. Namely, we adopt  
\begin{equation}\label{eq:L_nu}
L_{\omega}\propto \omega^{-a}\exp(-\omega/T_s)\,,
\end{equation}
where $T_s\sim {\cal O}(m_{\rm e})$ (we use 200 keV in the following, like in Ref.~\cite{Poulin:2017bwe}) and $|a|\lesssim 0.5$, again like in Ref.~\cite{Poulin:2017bwe} ($a=0$ was used in Ref.~\cite{Ali-Haimoud:2016mbv}). Before addressing the question of the effective PBH mass evolution, let us review typical benchmarks for $L_{\rm acc}$. Also, note that if the time-dependence in  $M(t)$ is not negligible, then $f_{\rm PBH}$ may become time-dependent as well. As we will argue below, this is not the case for the redshift range of interest. If comparing the CMB bounds with other, low-$z$ bounds, one should take into account that our $f_{\rm PBH}$ denotes the {\it initial} DM mass fraction in the form of PBH. 

\subsection{The  Eddington limit}
The Eddington luminosity $L_E$ is the luminosity at which accretion is balanced by radiation pressure  in a spherical system, and can be simply computed as
\begin{equation}
L_E\equiv \frac{4\pi\mu G M m_p\,c}{\sigma_T}\simeq 1.26\times 10^{38}\,\frac{M}{M_\odot}\,{\rm erg/s}\,,\label{Ledd}
\end{equation} 
with $\sigma_T$ the Thomson cross-section, $m_p$ the proton mass, and $\mu$ a chemical composition dependent parameter, equal to unity for pure hydrogen. 
Under the rather extreme hypothesis that $L_{\rm acc}=L_E$, we see from eq.~(\ref{eq:dQdt}) that the energy injected per unit volume per unit time scales like a constant times the matter dilution factor of the universe, so that the bounds should be independent of $M$.
Note that $L_E$ is a quantity linear in the accreting object mass, and allows one to introduce a fundamental accretion timescale (independent of the mass of the object) known as Salpeter time,
\begin{equation}
\tau_E\equiv \frac{M c^2}{L_E}=\frac{\sigma_T\,c}{4\pi\mu G m_p}\simeq 0.4\,{\rm Gyr}\,.\label{salpetertau}
\end{equation}
As we will illustrate more quantitatively in Sec.~\ref{SMBHprob}, under standard assumptions the PBH mass accretion timescale is about one tenth of eq.~(\ref{salpetertau}) for a BH shining at the Eddington luminosity. On the other hand, the age of the universe at the most relevant redshifts for CMB bounds, $ 300\lesssim z\lesssim 600$, is 1-3 Myr, more than a factor 100 smaller than eq.~(\ref{salpetertau}).  For less extreme accretion rates, the hierarchy between the accretion timescale and the age of the universe is {\it a fortiori} even bigger. As a result, the PBH mass can be considered constant and equal to its initial value at the epochs relevant for deriving CMB bounds.

\subsection{More realistic accretion scenarios}
The function $L_{\rm acc}$ in Eq.~(\ref{eq:dQdt}) is usually parameterized in terms of the two following quantities:
\begin{itemize}
\item[i)] $\dot M$, the matter accreted  per unit time onto the PBH. 
\item[ii)]$\epsilon$, the overall efficiency of conversion of accreted matter into radiation, in terms of which one writes:
\begin{equation}
L_{\rm acc}\equiv \epsilon \dot M\,c^2\,.\label{defepsilon}
\end{equation}
\end{itemize}

Concerning $\dot M$, an analytical theory exists in two limiting cases, both applying to a homogeneous gas of mass density $\rho_\infty$:
The hypothesis of a stationary, spherical symmetric accretion of a body at rest~\cite{Bondi:1952ni} (Bondi), and the purely ballistic limit (i.e. accounting only for gravity, no hydrodynamical nor thermodynamical effects included) of a point mass moving at a constant speed $v_{\rm rel}$ in the gas~\cite{1939PCPS...35..405H,1940PCPS...36..325H,1940PCPS...36..424H} (Hoyle-Lyttleton). These limiting cases justify the following ``Bondi-Hoyle-Lyttleton'' (BHL) parameterization
\begin{equation}
\dot{M}_{\rm BHL}= 4 \pi \lambda  \rho_{\infty} \frac{(G\,M)^2}{v_{\rm eff}^3}\,,\label{BHL0}
\end{equation}
where 
\begin{equation}
v_{\rm eff}^2\equiv v_{\rm rel}^2+c_s^2\,,
\end{equation}
$c_s$ is the speed of sound in the homogeneous matter of density $\rho_\infty$, and $\lambda$ is a dimensionless coefficient dependent upon environmental parameters, of ${\cal O}$(1) for $v_{\rm rel}\gg c_s$ (Hoyle-Lyttleton), and a calculable function assuming values of ${\cal O}$(0.1-1) (see e.g.~\cite{Ali-Haimoud:2016mbv}) in the limit $c_s\gg v_{\rm rel}$ (Bondi). For more realistic situations,  eq.~(\ref{BHL0}) is often used, but with $\lambda$ now intended as an adjustable parameter or function  fitted e.g. to simulation results. 
The cross-section for spherical accretion onto a pointlike particle is usually described in terms of the {\it Bondi radius}, which is the distance from the center at which the escape velocity equals the sound speed. Hence, it makes sense to  define a ``generalized'' Bondi radius,
\begin{equation}
r_{\rm B}\equiv \frac{GM}{v_{\rm eff}^2}\label{rBondigener}\,,
\end{equation}
in terms of which Eq.~(\ref{BHL0}) writes
\begin{equation}
\dot{M}_{\rm BHL}= 4 \pi \lambda  \rho_{\infty}\,v_{\rm eff}\,r_{\rm B}^2 \,,\label{BHL}
\end{equation}
and which, just like Eq.~(\ref{rBondigener}), reduces to Bondi's results for $v_{\rm eff}\to c_s$, while smoothly interpolating to the Hoyle-Lyttleton regime for larger velocities. 

Concerning $\epsilon$, it can be computed in spherical symmetry under some assumptions for the radiative processes, the most up-to-date treatment being provided in Ref.~\cite{Ali-Haimoud:2016mbv}. In that case, it assumes rather small values, of the order of $10^{-5}$. In the case of disk accretion, a typical benchmark value considered in the literature is $0.1$.
In the following, we adopt for $\lambda$ and $\epsilon$ the same prescriptions used in Ref.~\cite{Poulin:2017bwe}, and already implemented in the \texttt{ExoCLASS} package.

One of the main unknowns in the cosmological problem at hand is the actual relative velocity of PBH and baryons. The most obvious velocity scale in the problem is the sound speed in the baryon fluid, $c_s$. At large spatial scales, a larger velocity  $v_L$ is predicted in linear (but non-perturbative) theory~\cite{Tseliakhovich:2010bj}, but it is questionable if that applies down to the small-scales relevant for accretion, where the PBH potential dominates and the DM fluid approximation breaks down~\cite{Poulin:2017bwe}. Despite some first studies in that respect suggesting negligible effects in the epoch of interest~\cite{Hutsi:2019hlw}, gas simulations are definitely required to account for the dissipative nature of the baryonic gas accreting on the proto-halos. 
 We remain agnostic on the question, and consider two cases: 

i) If $v_L$ is a reasonable proxy, the PBH-baryon motion at the relevant redshifts is supersonic ($\cal{M}$ $\sim 2-5$) and, as argued for instance in Ref.~\cite{2013ApJ...767..163P,Ricotti17}, the accretion should then be disk-like~\footnote{ 
Note that this argument is independent of the angular momentum due to binaries or small-scale motion discussed in Ref.~\cite{Poulin:2017bwe}, hence applies also to small $f_{\rm PBH}$.}. In this case, our fiducial parameterization closely follows~\cite{Poulin:2017bwe}, with  
\begin{equation}
v_{\rm eff}\simeq \sqrt{c_{s}\sqrt{\langle v_{\rm L}^2\rangle}}\,,\label{veff}
 \end{equation}
 where, at $z\lesssim 1000$, we adopt~\cite{Ali-Haimoud:2016mbv,Tseliakhovich:2010bj}
 \begin{equation}\label{csvL}
 c_{{\rm s}}  \simeq 6 \frac{\rm km}{\rm s}\sqrt{\frac{1+z}{1000}}\,,\:\:\ \sqrt{{\langle v_{\rm L}^2\rangle}}\simeq 30\frac{\rm km}{\rm s}\, \left( \frac{1+z}{1000}\right)\,.
 \end{equation}

ii) If, on the other hand, the relevant small-scale motions are subsonic ( i.e. $\cal{M}$ $\lesssim 1$), for the low values of $f_{\rm PBH}$ considered in the following the accretion should be better approximated by a spherical one. In this case, a conservative, spherical accretion scenario is adopted, with Bondi accretion and  $v_{\rm eff}\simeq c_s$, with $c_s(z)$ obeying Eq.~(\ref{csvL}). We also assume the limiting case of  purely collisional ionization, known to yield the most conservative bounds~\cite{Ali-Haimoud:2016mbv}.

\subsection{Regimes of validity}
The treatment just described has two limitations:

i) The hypothesis of {\it stationarity}, i.e. the system settles down in the Bondi steady-state fast compared to the cosmological expansion:
\begin{equation}
\frac{r_{\rm B}}{v_{\rm eff}}H(z)<1\,.
\end{equation}
As already argued in Ref.~\cite{Ricotti:2007jk}, this leads to the requirement $M\lesssim {\rm few}\times 10^4\,M_\odot$, which is the upper limit for which we present our results.

ii) How to deal with super-Eddington accretion, i.e. $L_{\rm acc}>L_E$, which in our formalism can be attained for sufficiently heavy PBH.  It is still debated what happens under these accretion conditions, which are sensitive to multi-dimensional effects and realistic disk radiation spectra (for a recent study, see~\cite{Takeo:2019uef}). It is clear however that several phenomena come into play:  For instance, radiative and kinetic feedback can break stationary conditions, with outflows and episodic periods of very high luminosity alternating with long period of low accretion and luminosity. Or quasi-steady state, super-Eddington {\it mass} accretion can take place, with a corresponding drop in efficiency in order to satisfy $L_{\rm acc}\lesssim L_E$. While we will comment again on this regime in Sec.~\ref{SMBH}, we address the reader to reviews such as Ref.~\cite{Mayer:2018vrr} for details and a more complete picture.
In the following, we will adopt the prescription to cap luminosity at $L_E$ whenever the formalism yields nominally $L_{\rm acc}>L_E$. 

Interestingly enough, super-Eddington accretion is attained in our formalism for $M\gtrsim 10^4\,M_\odot$, so that both conditions above yield similar limitations, albeit by coincidence. The homogeneous approximation discussed in Appendix~\ref{ionsizecheck} is also valid in the same range of interest.  It is worth clarifying that CMB anisotropy bounds are expected to exist also at higher masses, but they become rather uncertain and definitely the formalism above is insufficient to tackle them. 
Fortunately, for such high masses, other bounds  become relevant, as discussed in Sec.~\ref{primSMBH}.

In App.~\ref{friction}, we also check that the {\it dynamical friction} that a rather massive PBH experiences moving supersonically in the cosmological baryonic gas is negligible for the masses and redshifts of interest for this work.

\section{Including cosmological DM halos}\label{loophole}
It has been argued in the past~\cite{Bertschinger:1985pd,Mack:2006gz,Berezinsky:2013fxa} that,  due to the PBH gravity, a DM halo would form around massive PBH, boosting their accretion. 
Note that the Eddington luminosity benchmark only applies to baryons, subject to radiation pressure. As far as baryonic accretion is the only one considered, the Salpeter timescale suggests that the PBH mass remains essentially constant down to the redshifts of interest for CMB bounds. Hence, we can safely consider $f_{\rm PBH}$ constant, while DM halos affect the phenomenology via the altered accretion rate. 

Although the original Bondi problem was considering accretion onto a point particle, a natural generalization of the notion of Bondi radius for an extended distribution of mass can be written as~\cite{Park:2015yrb}:
\begin{equation} \frac{G_N\,M_{\rm PBH}}{r_{\rm B,eff}}-\Phi_h(M_{\rm PBH},r_{B,{\rm eff}},t)=v_{\rm eff}^2(t)\,,\label{effRB}
\end{equation}
where $r_{\rm B,eff}$, the effective Bondi radius, is the unknown,  $M_{\rm PBH}$ is the initial PBH mass and $\Phi_h$ the (time-dependent) gravitational potential associated to the DM halo.
Our treatment of the problem consists in adopting Eq.~(\ref{BHL}), but with $r_{\rm B}$ replaced by $r_{\rm B,eff}$, solution of Eq.~(\ref{effRB}) with the gravitational potential of the halo estimated analytically or numerically.  In this section, we revisit our estimates accounting for the DM capture phenomenon. First, we consider  a toy model, which we believe determines an upper limit to the effect. Then, we improve over this estimate with the help of numerical simulations.

\subsection{Toy model}
\begin{figure}[htbp]
\begin{center}
\includegraphics[width=0.45\textwidth]{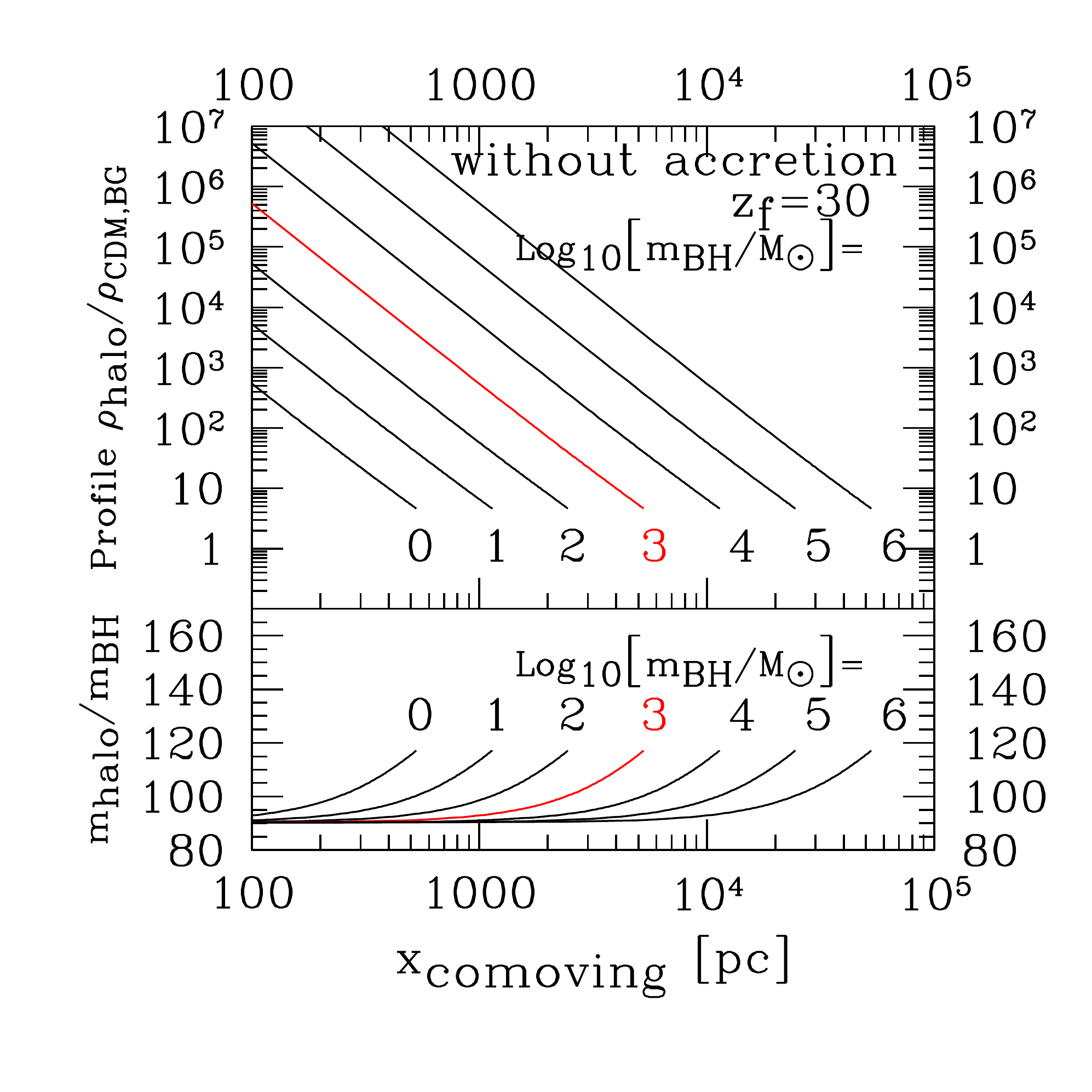}
\vspace{-0.1cm}
\caption{Density profile of halo around a PBH and integrated
  halo mass as a function of comoving radius. We plot the cases at
  $z = 30$ for
  $M_{\rm PBH} = 1, 10, 10^2, 10^3, 10^4, 10^5, 10^6 M_{\odot}$.}
\label{fig:xrhom}
\end{center}
\end{figure}

The most optimistic scenario for PBH growth is that DM is exactly cold and with no dispersion, and the PBH is the only center of attraction in the whole universe. This is a spherically symmetric problem. 
In order to calculate the time evolution of a radius $r$ of a mass-shell around
a PBH which encloses different species, we solve the following differential equation,
\begin{eqnarray}
  \label{eq:d2rdt21}
  \frac{d^2r}{dt^2} =-\frac{4 G_{\rm N}} \pi{3}
  r \left[\rho_{\rm PBH}+  \sum_i 
\left( \rho_i + 3 p_i \right)\right],
\end{eqnarray}
where $\rho_i$ and $p_i$ are the energy density and pressure of a component ``i'', respectively, and we defined the energy density of the PBH as
$\rho_{\rm PBH} = 3\,M_{\rm PBH}/(4 \pi r^3)$. Here $i$ runs on all the components of radiation and matter (and dark energy if it were effective).  The
physical radius $r$ is represented by $r= a(t)x$ where $a(t)$ is the
scale factor normalized to be $a(t_0) = 1$ at the present Universe
($t=t_0$), and $x$ is the co-moving coordinate. 
At each time $t$, the bound (or halo) mass is equivalent to the DM density up to the radius $r_s$ defined by 
\begin{equation}
\frac{d r_s}{d t}(t)=0\,, \label{halocond}
\end{equation}
although a similar value would be found if one derives $r_s$ from the condition that the density at $r_s$ is twice the cosmological background one, as in Ref.~\cite{Mack:2006gz}. Under the above-mentioned approximations, we find good agreement with the results reported in Ref.~\cite{Mack:2006gz}, namely:
\begin{itemize}
    \item A time evolution given by
\begin{eqnarray}
  \label{eq:scalingLawCorr}
  M_{\rm halo} \simeq \left( \frac{3000}{1 + z} \right)  M_{\rm PBH}\,.
\end{eqnarray}
\item A density profile proportional to $\propto r^{-3}$, as illustrated in Fig~\ref{fig:xrhom}, down to the distances (not resolved in~Fig~\ref{fig:xrhom}) where a free-fall profile $r^{-3/2}$ takes over.
\end{itemize} 

Eq.~(\ref{eq:scalingLawCorr}) should be understood as an upper limit to the mass growth of the PBH via DM accretion proceeding self-similarly once a DM halo of mass larger than the PBH is accumulated. Its breakdown is only expected at very late times (e.g. when dark energy kicks in) or when the hypothesis of isolated PBH breaks down (which further requires high $f_{\rm PBH})$. On the other hand, the radial profile crucially depends on 
the free-fall boundary condition at the center. Not accounting for the DM  angular momentum is however a very  crude approximation. It was speculated in Ref.~\cite{Ricotti:2007jk}, Sec. 4, that other scaling solutions like the ones described in the seminal paper by Bertschinger~\cite{Bertschinger:1985pd} may provide a better description of the results. To verify this conjecture, we turn to the results of N-body simulations.

\subsection{Numerical simulations}

We perform cosmological $N$-body simulations using a version of the {\sc CUBEP$^3$M} code \citep{HarnoisDeraps:2012vd} modified to include PBHs as a separate particle species co-evolved with a generic collisionless DM candidate \citep{Inman:2019wvr}.  We opt for homogeneous DM initial conditions at $a=10^{-6}$ and select cosmological parameters consistent with Planck: $\Omega_c=0.26$, $\Omega_b=0.05$ and $z_{\rm eq}=3374$, with $\Omega_{\rm PBH}=f_{\rm PBH}\Omega_c$.  Baryons are not evolved and instead assumed to be homogeneous inducing errors of order $\Omega_b/\Omega_m\simeq 15\%$.  With this setup, simulations are invariant to the numerical masses of the particles, instead being sensitive to their ratio:  
\begin{align}
\frac{M_{\rm PBH}}{M_{\rm DM}} = \frac{f_{\rm PBH}}{1-f_{\rm PBH}} \frac{N_{\rm DM}}{N_{\rm PBH}}.
\end{align}
After running a simulation, we can set $M_{\rm PBH}$ to a physical value which then fixes the volume, $L^3$, of the simulation:
\begin{align}
\label{eq:mpbh}
M_{\rm PBH} = \bar{\rho}_{\rm cr} \Omega_c f_{\rm PBH} \frac{L^3}{N_{\rm PBH}}\,,
\end{align}
where $\bar{\rho}_{\rm cr}$ is the comoving critical density.

We would like our simulations to have the best possible length resolution (i.e.~the smallest box size $L$).  From Eq.~(\ref{eq:mpbh}) we see that, at fixed $M_{\rm PBH}$ and cosmological parameters, this is achieved by minimizing the number of PBHs in the simulation, i.e.~$N_{\rm PBH}=1$.  By doing this we no longer accurately follow Poisson fluctuations in the PBH density field; however, from the Epstein mass function \citep{10.1093/mnras/205.1.207,1995MNRAS.276..796S} describing the Poisson distribution we can deduce that PBHs rarely interact when $f_{\rm PBH} \ll (1+z) \times 10^{-4}$.  We performed an explicit test of this by running a simulation with $f_{\rm PBH}=10^{-5}$ and $N_{\rm PBH}=100$ and found the resulting profiles comparable, but noisier, to the single PBH case.  To accurately model the isolated halo growth we require that the DM halo be composed of many DM particles when it becomes comparable to $M_{\rm PBH}$:
\begin{align}
f_{\rm PBH} \gg \frac{1}{1+N_{\rm DM}} \simeq N_{\rm DM}^{-1} \simeq 4\times10^{-9}
\end{align}
where the last equality utilises our maximum number of particles: $2\times512^3$.  We therefore consider $f_{\rm PBH}=10^{-5}$, $10^{-6}$ and $10^{-7}$ with $N_{\rm DM}=2\times384^3$, $2\times512^3$ and $2\times512^3$.  
The simulations are run from $a=10^{-6}$ to $a=10^{-2}$. 
We have tested how much halting the energy injection at this redshift affects our constraints and found it to be around a percent. 

\begin{figure}[h]
\begin{center}
\includegraphics[width=0.45\textwidth]{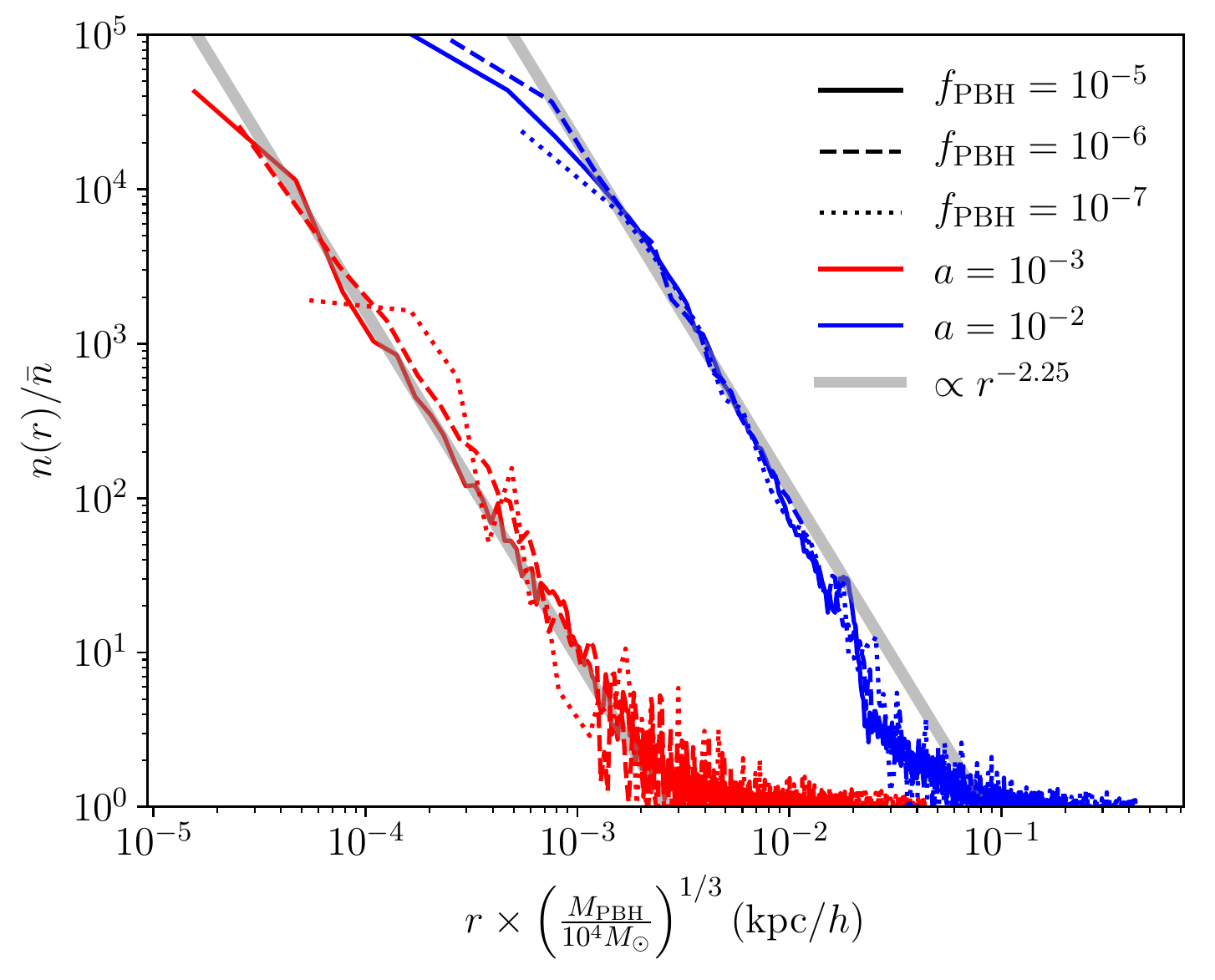}
\caption{DM halo profiles (with mass scaling in x-axis) around a PBH at $z=99$ and $z=999$, for different values of the PBH abundance. A power-law profile $r^{-2.25}$ is also shown for comparison.}
\label{fig:dNdVn}
\end{center}
\end{figure}

Given our assumption that the PBHs are isolated, we expect the DM halo to be independent of $f_{\rm PBH}$ apart for numerical resolution effects.  We show the density profile as a function of $f_{\rm PBH}$ at $a=10^{-3}$ and $10^{-2}$ in Fig.~\ref{fig:dNdVn}.  As expected, we find that the profiles are independent of $f_{\rm PBH}$.  We also find that the best resolution is obtained for the $f_{\rm PBH}=10^{-5}$ simulation, despite the fact that it has slightly fewer particles.  We therefore use the results from this simulation.  We also see the effects of redshift on the profile.  Over almost two decades in radius, the profile at early times matches the $r^{-2.25}$ power-law predicted by~\cite{Bertschinger:1985pd}; at late times, it is only slightly steeper, moving closer to $r^{-2.5}$.  This power law profile is consistent with those found in (independent) numerical simulations by \citet{Adamek:2019gns}, who also find a smooth transition to standard NFW-like profile at large radii.

\begin{figure}[h]
\begin{center}
\includegraphics[width=0.45\textwidth]{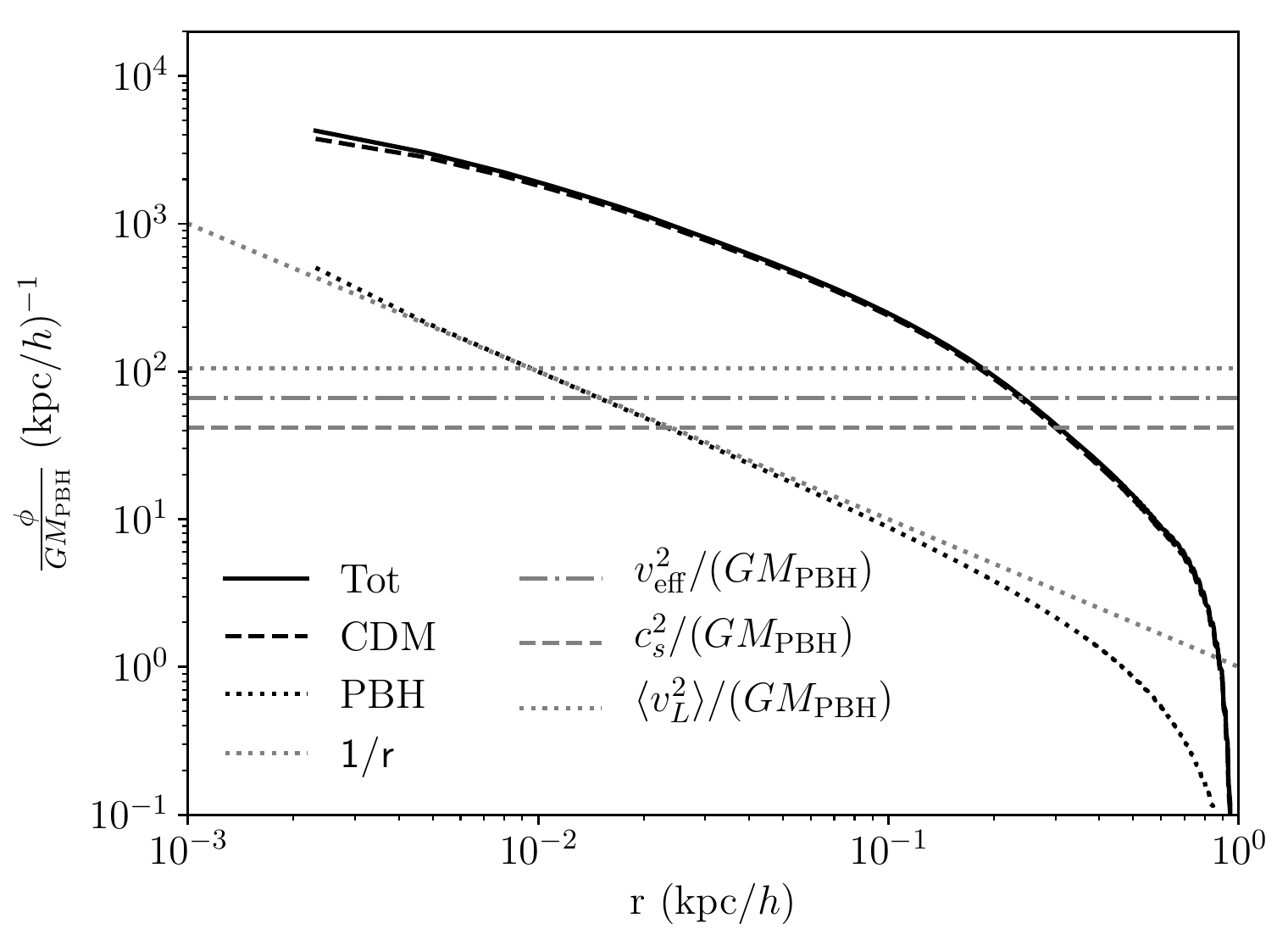}
\caption{Gravitational potentials of the PBH (dotted inclined lines, with the straight line representing the analytical result), the halo (curved dashed line) and the sum (solid line) compared to the typical velocity scales of the problem (horizontal lines) at $z=99$.}
\label{fig:phiGM}
\end{center}
\end{figure}

Our chief goal is determining $r_{\rm B,eff}$.  To do this, we first interpolate the particles to a grid using the Cloud-in-Cell method and then solve Poisson's equation for the gravitational potential:
\begin{align}
\phi_{i} = (4\pi G)^{-1} \nabla^{-2} \rho_i
\end{align}
where $_i$ can indicate PBH and CDM separately and $\nabla^{-2} \rho$ is evaluated in Fourier space.  We then find $r_{\rm B,eff}$ by plugging the potential above in Eq.~(\ref{effRB}).
We show the obtained potentials at $z=99$ in Fig.~\ref{fig:phiGM}, with $r_{\rm B,eff}$ defined as the intersection of the total potential (solid black) with the horizontal grey line.  We remark some numerical artifacts: For the PBH potential we know the exact result, $-\phi_{\rm PBH}(r)/(G M_{\rm PBH})=1/r$.  However, we see that the result differs from this analytical result on both small and large scales:  On small scales this is due to the interpolation error, whereas on large scales it is due to periodic boundary conditions.
Typically, the PBH contribution to the halo is negligible whenever the DM distribution is important. The periodicity artifact leads to a slight underestimates of $r_{\rm B,eff}$. At high redshifts the PBH is much more relevant, $r_{\rm B,eff}$ is smaller and the numerical solution may lead to a slight overestimate of the solution. 
It is interesting to note that the numerical results lead to an estimated halo mass which about 60\% of the simple result of Eq.~\ref{eq:scalingLawCorr}, with a similar scaling with redshift, although with a different mass profile. These results motivate the semi-analytical model described in the following section, which we later use to cover PBH masses whose Bondi radii are not resolved by our simulations.

 Note that it is safe to neglect the ``ordinary'' DM halos feedback onto the halos growing around PBH, since the former ones only grow at much later times (typically $z\lesssim 30$ in a $\Lambda$CDM cosmology) than those of concern for us. {\it A fortiori}, the feedback of the baryons can also be neglected. The bulk of baryons has significant velocity at the epochs of interest, and they are still kinetically coupled to the CMB. Most of them are essentially unbound to halos, and their ratio to the DM in the growing halos around PBH is much smaller than the baryon to dark matter cosmological density ratio of $\sim 15\%$. Hence, objections on the realism of power law DM density profiles  around BH surviving in the current universe~\cite{Ullio:2001fb} do not apply to the pristine configurations considered here.

\subsection{Semi-analytical model}\label{procedure}
 In the specific case of a point-like potential due to the PBH plus  the power-law matter distribution around it, with density $\rho(r)\propto r^{-\alpha}$ up to a distance $r_h$ and total mass $M_h$, Eq.~(\ref{effRB}) rewrites
\begin{eqnarray} v_{\rm eff}^2(z)=
\frac{G_N\,M_{\rm PBH}}{r_{B,{\rm eff}}}+\frac{G_N M_h}{r_{B,{\rm eff}}}\Big{\{}\Theta(r_{B,{\rm eff}}-r_h)+\nonumber\\
\left.+\frac{\Theta(r_h-r_{B,{\rm eff}})}{1-p}\left[\left(\frac{r_b}{r_h}\right)^p-p\left(\frac{r_{B,{\rm eff}}}{r_h}\right)\right]\right\}\,,\label{effbondi}
\end{eqnarray}
where $p=3-\alpha$, and $M_h$ and $r_h$ depend from $\{M_{\rm PBH},z\}$.

We adopt Eq.~(\ref{eq:scalingLawCorr}) for  the halo mass within the turnaround radius,
where the turnaround radius is (see e.g. Sec. 4 in Ref.~\cite{Ricotti:2007jk})
\begin{equation} 
r_{\rm t.a.}\simeq 58\, {\rm pc}\,(1+z)^{-1}\left(\frac{M_h(M_{\rm PBH},z)}{M_\odot}\right)^{1/3}\,.
\end{equation}
We identify $r_h=r_{\rm t.a.}$, in order to have a self-consistent normalization of the mass.

Eq.~(\ref{effbondi}) admits either the solution
\begin{equation}\label{smallhalo}
r_{\rm B,eff}=\frac{G_N(M_{\rm PBH}+M_{h})}{v_{\rm eff}^2}\simeq  \frac{G_N\,M_{h}}{v_{\rm eff}^2}\equiv r
_{{\rm B},h}\,,
\end{equation}
which holds if $r_h<r_{{\rm B},h}$; otherwise, if $r_h> r_{{\rm B},h}$, neglecting the PBH mass one has
\begin{equation}\label{largehalo}
r_{\rm B,eff}\simeq r_h
\left[(1-p)\frac{r_h}{r_{{\rm B},h}}+p\right]^{\frac{1}{p-1}}\,\leq r_h.
\end{equation}
Note that Eq.~(\ref{largehalo}) tends to $r_{{\rm B},h}$ when $p\to 0$, as expected: When the DM halo profile is very steep and/or the halo is very compact, as far as accreting baryons are concerned they simply see a BH whose effective mass is the sum of the PBH and the DM halo mass. If the halo is fluffy or large, only a fraction of the mass of the halo contributes to the accretion.
In any case, the condition $r_{{\rm B, eff}}\geq r_{\rm B, PBH}$ must hold. This constraint must be verified and eventually imposed by hand as a lower limit if using the approximated Eq.~(\ref{largehalo}) or the RHS of Eq.~(\ref{smallhalo}). 
We have found that the CMB constraints obtained using this model are in agreement within $50\%$ with the ones obtained from results of the numerical simulations in the mass range covered by the simulations\footnote{We checked that varying $p \in [0.50,0.75]$ affects our results to below 10\% level.}. We thus use this model  with $p=0.75$ to compute the impact of PBH accretion onto the CMB.

\section{CMB Constraints}\label{constraints}

\subsection{Impact of accretion onto DM halos}
\begin{figure}
\includegraphics[scale=0.4]{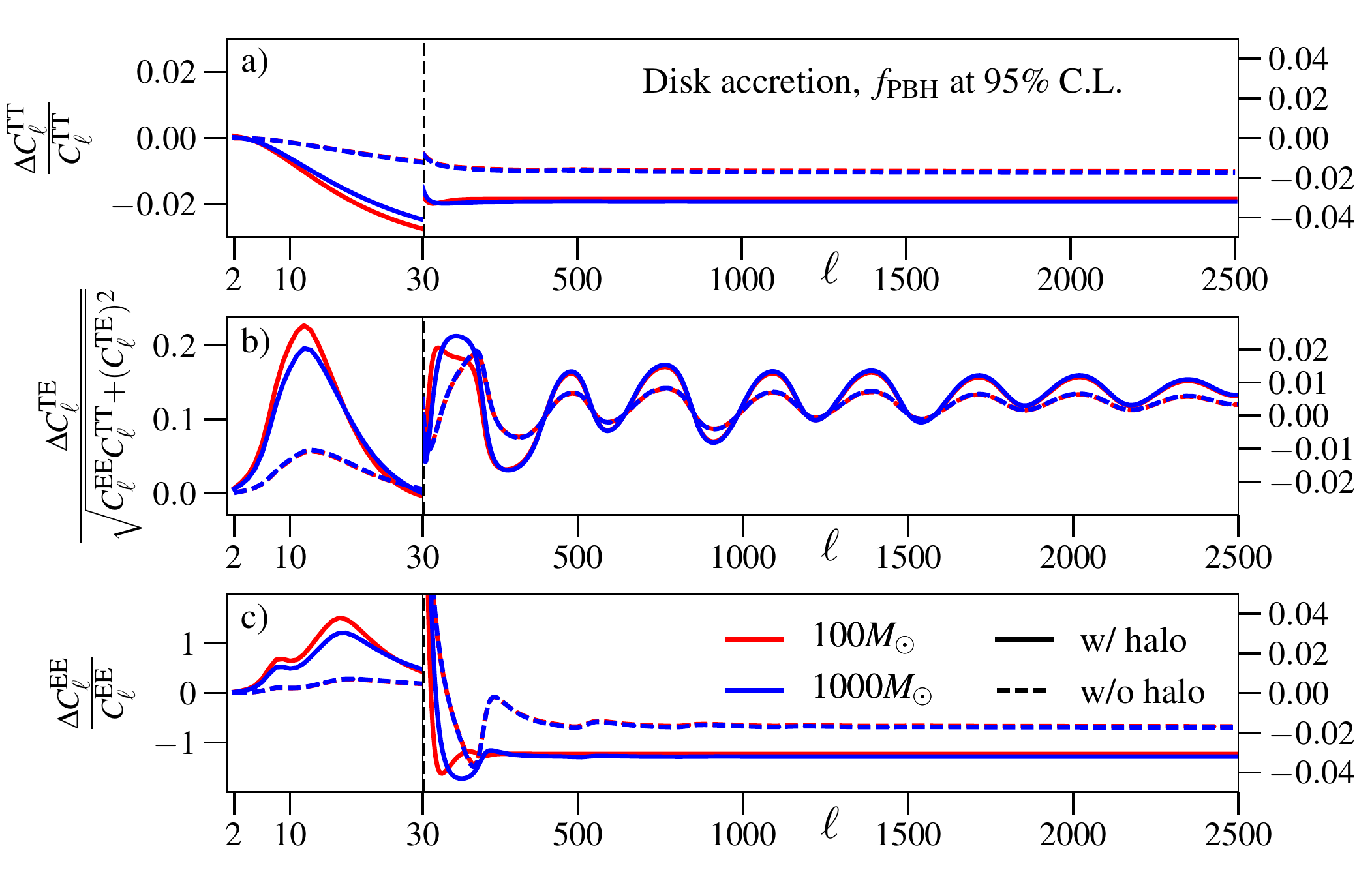}
\caption{Impact of accreting PBH with and without DM halos on the lensed CMB TT (panel a), TE (panel b) and EE (panel c) power spectra. The reference is $\Lambda$CDM with $\{\omega_b=0.02218,\omega_{\rm cdm} = 0.1205, 100*\theta_s=1.04069,\tau_{\rm reio}=0.055, {\rm ln}(10^{10}A_s)=3.056,n_s=0.9619\}$. We consider a disk accretion scenario and set the PBH fraction to the constraints at 95\% C.L. derived in this work. }
\label{fig:cmb}
\end{figure}

\begin{figure}
\includegraphics[scale=0.4]{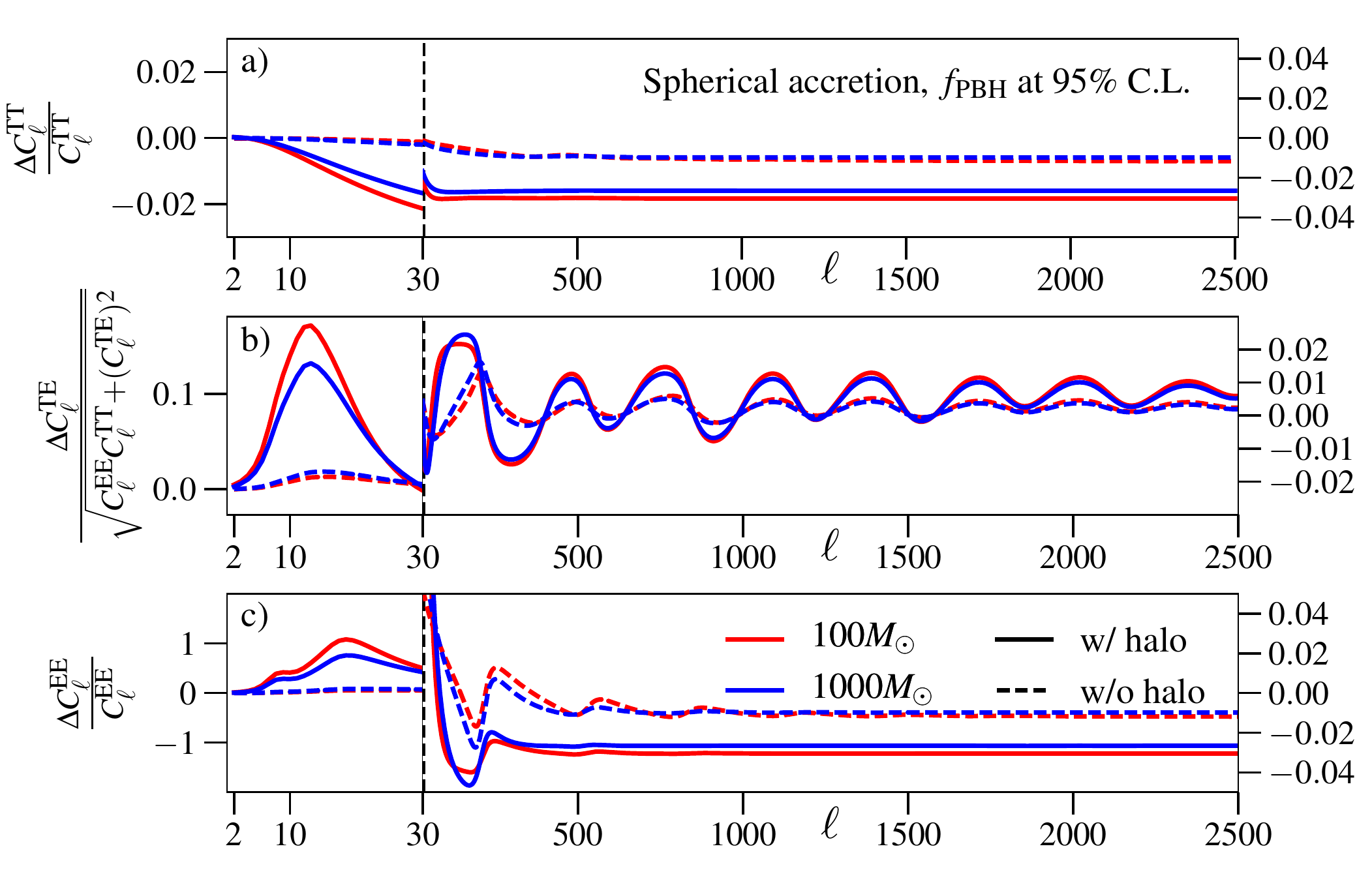}
\caption{Same as Fig.~\ref{fig:cmb}, in the spherical accretion case.}
\label{fig:cmb2}
\end{figure}

We modify the branch \texttt{ExoCLASS} \cite{Stocker:2018avm} of the public code \texttt{CLASS} to include the effect of DM halo. We compare the effect of accretion with and without halos. 
In practice, our simulations only have the necessary resolution to solve eq.~(\ref{effRB}) for redshifts $100<1+z<1000$, which encompasses the redshift range at which e.m. energy deposition has the biggest impact on the CMB, $300<1+z<600$ \cite{Slatyer09}. We extrapolate with constant values of the lower (higher) boundary at lower (higher) redshifts. We checked that this has sub-percent impact by turning off injection at $z<100$, while the effect of energy ejection is naturally turned off at higher$-z$ since the plasma is still mostly ionized. We show the effect of the e.m. energy injection from accretion of matter around PBH on the CMB power spectra in Figs.~\ref{fig:cmb} and \ref{fig:cmb2} for $M_{\rm PBH}/M_\odot = 100,1000$. Since we find that including the halos increase the impact of the PBH on the power spectra by up to $\sim 2$ orders of magnitude (at fixed fraction), we actually compare two values of $f_{\rm PBH}$ such that each case shows the 95\% C.L. exclusion. Interestingly, because of different time-dependence of the energy injection, the shape of the EE and TE CMB power spectra residuals is significantly different (see e.g. Ref.~\cite{Poulin:2017bwe} for a review of the effect of e.m. energy injection). This shows that the effect of DM halos could potentially be distinguished (and thus should be taken into account) if a signal were detected. It might even tell us something about the nature of DM as halos may not form depending on DM properties (e.g. if DM is warm or fuzzy). Further work will be required to accurately characterize the signal from PBH accretion given the sensitivity of future CMB experiments to polarization anisotropy.

\subsection{Analysis}

We run a Markov-chain Monte Carlo (MCMC) using the public code \texttt{MontePython-v3}\footnote{\url{https://github.com/brinckmann/montepython_public}} \citep{Audren:2012wb,Brinckmann:2018cvx}, interfaced with our modified version of \texttt{CLASS}.
We perform the analysis with a Metropolis-Hasting algorithm, assuming flat priors on $\{\omega_b,\omega_{\rm cdm},\theta_s,A_s,n_s,\tau_{\rm reio},f_{\rm PBH}\}$ at fixed PBH mass $M/M_\odot = [10,20, 30,40,50,10^2,10^3,10^4,10^5]$. We additionally perform a MCMC run at fixed $f_{\rm PBH} = 1$, with $M/M_\odot$ free to vary to determine the minimal PBH mass probed by cosmological data through accretion. In practice we find that $M_{\rm min}^{95\%} = 15 M_\odot >10M_\odot$ in the spherical case, we therefore did not run with $M=10M_\odot$ in that accretion scenario. We adopt the {\em Planck} collaboration convention and model free-streaming neutrinos as two massless species and one massive with $M_\nu=0.06$ eV. 
Our data set includes {\em Planck} 2018 high-$\ell$ and low-$\ell$ TT, EE and lensing likelihood \cite{Aghanim:2018eyx,Aghanim:2019ame}; the isotropic BAO measurements from 6dFGS at $z = 0.106$~\cite{Beutler:2011hx} and from the MGS galaxy sample of SDSS at $z = 0.15$~\cite{Ross:2014qpa}; the anisotropic BAO and the growth function $f\sigma_8(z)$ measurements from the CMASS and LOWZ galaxy samples of BOSS DR12 at $z = 0.38$, $0.51$, and $0.61$~\cite{Alam:2016hwk}. Additionally, we use the Pantheon\footnote{\url{https://github.com/dscolnic/Pantheon}} supernovae dataset \cite{Scolnic:2017caz}, which includes measurements of the luminosity distances of 1048 SNe Ia in the redshift range $0.01 < z < 2.3$. As usual, we use a Choleski decomposition \citep{Lewis:2013hha} to deal with the numerous nuisance parameters associated with the likelihoods (not recalled here for brevity). We consider chains to be converged using the Gelman-Rubin \citep{Gelman:1992zz} criterion $R -1<0.05$. We perform four sets of runs, assuming either spherical or disk accretion, and absence or presence of a DM halo around the PBHs. For simplicity, we adopt a monochromatic PBH mass function, keeping in mind that for extended mass functions (which are to be generically expected from single-field inflationary models~\cite{Byrnes:2018txb}) bounds typically tighten~\cite{Kuhnel:2017pwq,Carr:2017jsz}, as we explicitly checked for the CMB ones in our previous article~\cite{Poulin:2017bwe}. 

\begin{figure}
\includegraphics[scale=0.33]{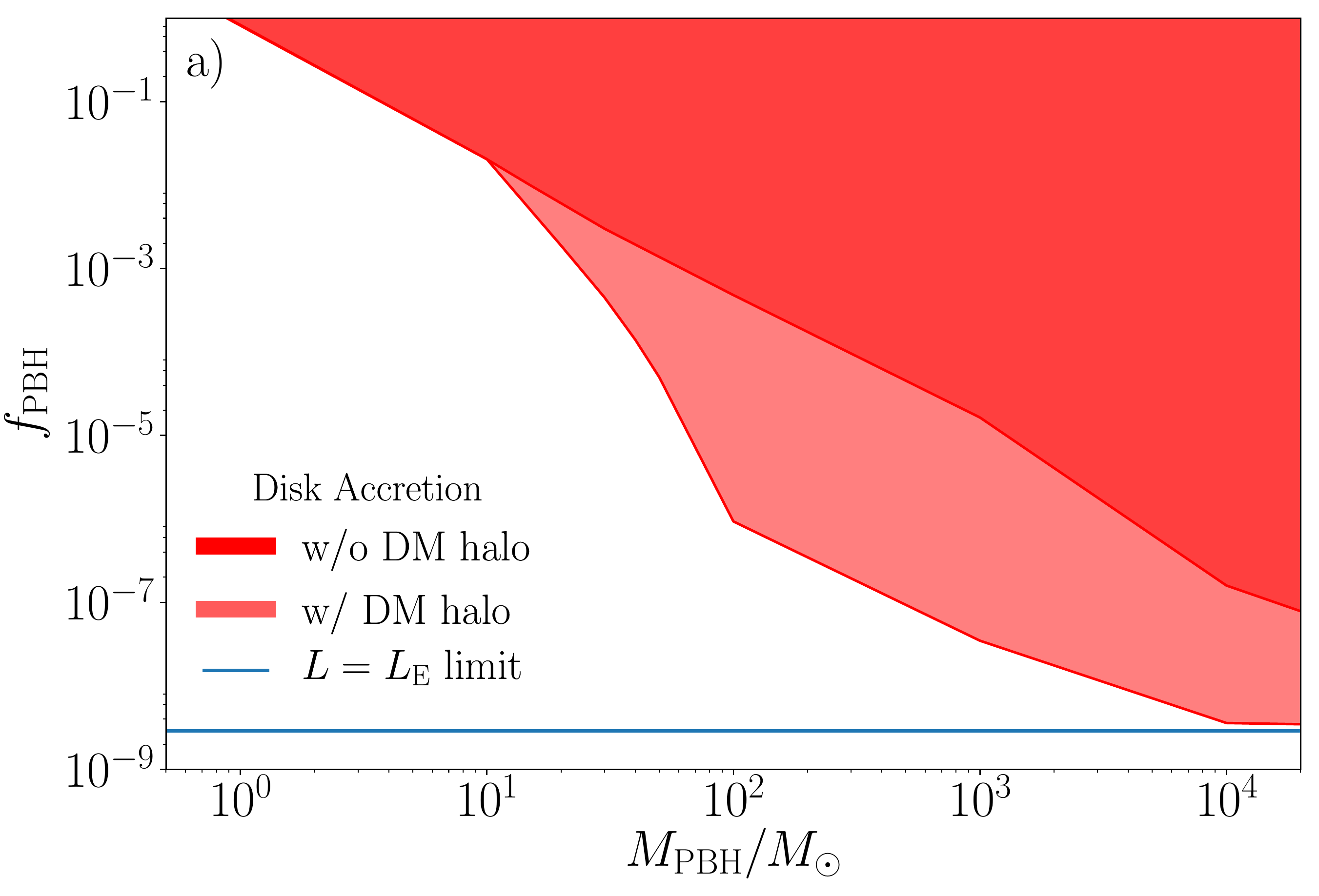}
\includegraphics[scale=0.33]{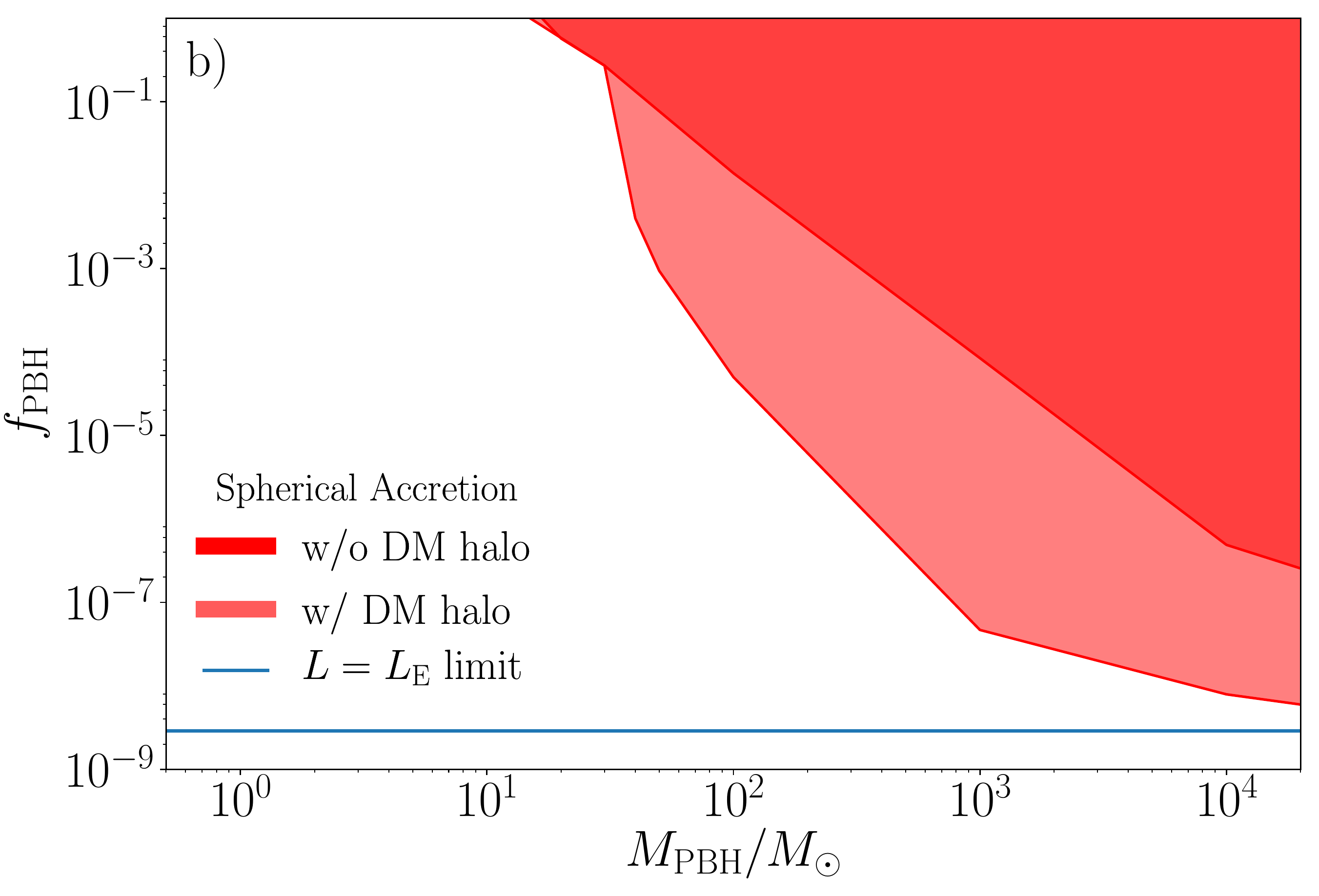}
\caption{Bounds on the abundance of PBH assuming disk accretion (panel a) or spherical accretion (panel b). We show the results with (light-shaded) and without (dark-shaded) the formation of a DM halo. The horizontal line shows the limiting bound of Eq.~(\ref{LEconstraint}). See text for details.}
\label{fig:bounds}
\end{figure}

As a warm-up, we derive the limit in case PBH are accreting at Eddington luminosity. As argued, in this case one obtains a mass-independent bound, which reads
\begin{equation}
f_{\rm PBH}<2.9\times10^{-9}\:\:\:(L_{\rm acc}=L_E).\label{LEconstraint}
\end{equation}
This is an optimistic benchmark for what is presumably the best limit that CMB can yield to. 
Our more realistic constraints at 95\% C.L. are shown in Fig.~\ref{fig:bounds}.  Note that the bounds on PBH in absence of DM halos (dark shaded regions) are themselves stronger than bounds previously derived in~\cite{Poulin:2017bwe} by a factor $\sim 4$. This improvement is due roughly equally to the new {\em Planck} 2018 low multipole polarization data and the additional use of BAO and Pantheon data, as well as to the improvements in the treatment of energy deposition, now implemented in \texttt{ExoCLASS}. Accounting for the the halo (light shaded regions) does not lead to significant differences unless $f_{\rm PBH}$ is sufficiently small, i.e. there is sufficient material for growing a sizable DM halo. The threshold to see significant improvements is $f_{\rm PBH}\lesssim 0.01$ for the disk accretion case, but already at $f_{\rm PBH}\lesssim 0.2$ for the spherical accretion case. For the latter case, the steep improvement of the bound around $M\sim 30\,M_\odot$ in presence of a halo is only indicative, since for $f_{\rm PBH}\gtrsim 0.01$ a non-negligible fraction of the DM can be gravitationally bound to two or more PBH, and the radial profile derived in the isolated-PBH approximation breaks down~\cite{Inman:2019wvr}. At higher masses,  sensitive to lower $f_{\rm PBH}$, the approximation is however robust: The formation of a DM halo around the PBH can strikingly improve the bound by up to $\sim 2$ orders of magnitude in the covered mass range. 
Our results also show that the bounds eventually flatten when $M\gtrsim 10^4\,M_\odot$. This is a consequence of the accretion attaining the Eddington limit for longer and longer periods of time, thus converging to Eq.~(\ref{LEconstraint}). As previously argued, in this range the bounds become shaky since the working hypotheses break-down.

\begin{figure*}[t!]
\centering
\includegraphics[scale=0.6]{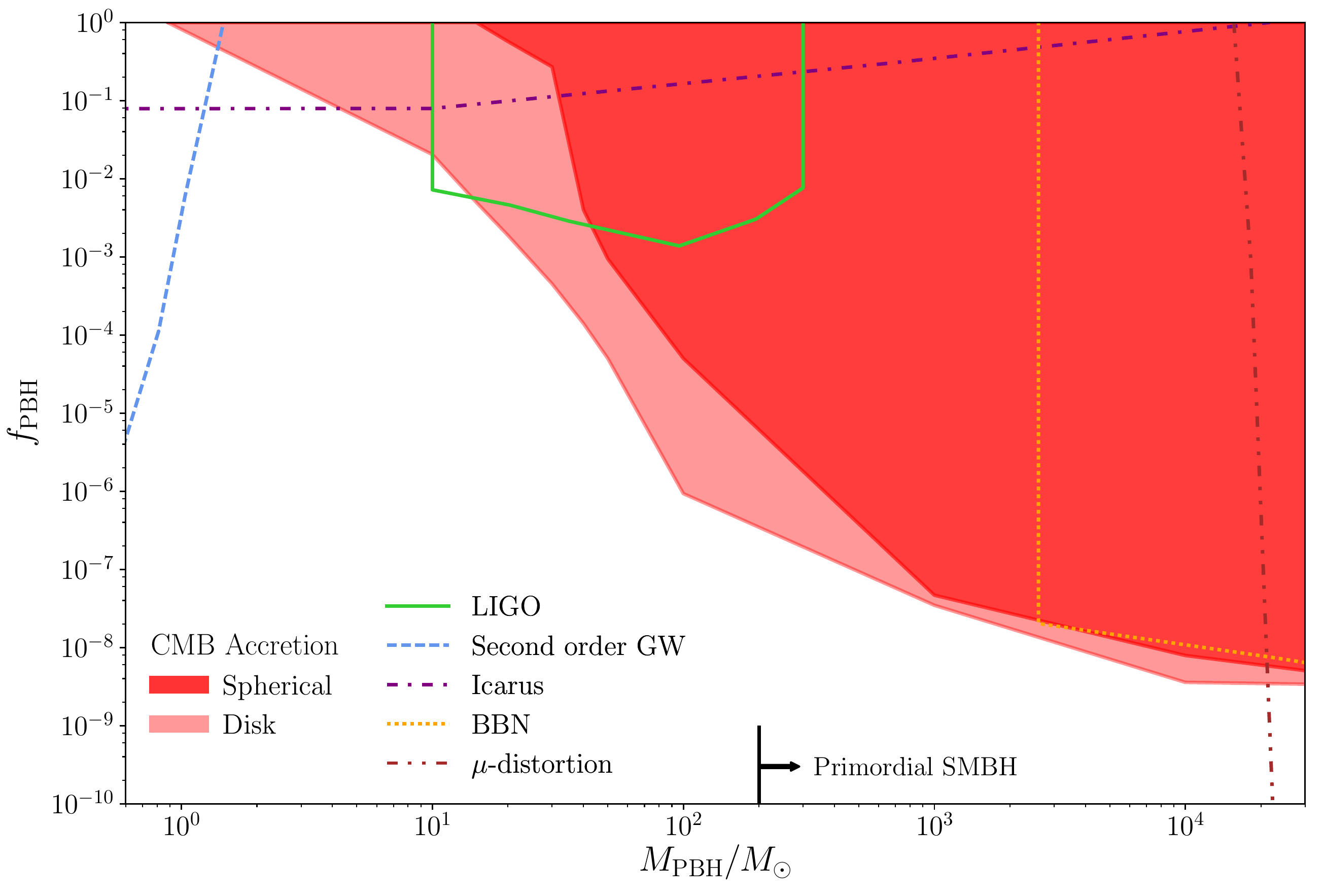}
\caption{Bounds on the abundance of PBH assuming disk accretion (light shade) or spherical accretion (dark shade), accounting for the formation of a DM halo. The most stringent bounds in the same mass regions are also shown: Second order GW ones from~\cite{Carr:2020gox}, Icarus ones from~\cite{Oguri:2017ock},  LIGO ones according to~\cite{Ali-Haimoud:2017rtz}, BBN bounds from~\cite{Inomata:2016uip}, spectral CMB distortions from \cite{Kohri:2014lza}. The arrow indicates that for masses $M\gtrsim 200\,M_\odot$, PBH can in principle grow in mass up to $10^9\,M_\odot$ by $z=7.5$ by accreting baryons at Eddington luminosity with $\epsilon=0.1$.}
\label{fig:summary}
\end{figure*}

It is worth commenting on the relative strength of the derived bounds with other existing ones, with the most stringent ones reported in Fig.~\ref{fig:summary}. Since 
curvature perturbations couple to tensor perturbations at second-order, PBH below the solar-mass scale are associated to GWs  generated in conjunction with their formation, falling in the frequency probed by pulsar timing arrays. The non observation of a stochastic signal in the nHz range sets tight bounds~\cite{Chen:2019xse,Carr:2020gox}. Galactic microlensing constraints~\cite{Tisserand:2006zx,Wyrzykowski:2011tr,Green:2016xgy}, roughly excluding  $f_{\rm PBH}\gtrsim {\cal O}(0.1)$, also apply. Other ``direct'' bounds come from the non-observation of mergers by LIGO/Virgo~\cite{Abbott:2018oah}. At few solar masses, leading constraints come from caustic crossing events in giant arcs (produced by stars embedded in high magnification regions due to a Galaxy cluster)~\cite{Oguri:2017ock}, but one may expect similar or tighter constraints from the extrapolation of the analysis of Ref.~\cite{Abbott:2018oah} to higher masses.

In the 10-100 $M_\odot$ range, the binary coalescence rate inferred by LIGO/Virgo is estimated to yield bounds at a level between $10^{-3}$ and $10^{-2}$~\cite{Sasaki:2016jop,Ali-Haimoud:2017rtz}, a result whose robustness to a number of effects has been checked in~\cite{Raidal:2017mfl,Kavanagh:2018ggo,Ballesteros:2018swv}. Note however that, according to~\cite{Raidal:2018bbj,Vaskonen:2019jpv}, accounting for binary disruption can relax  these limits to some extent.

 Other constraints at $M\sim {\cal O}(10)\,M_\odot$ roughly in the ballpark of $f_{\rm PBH}\lesssim {\cal O}(0.1)$ come from the non-observation of a stochastic gravitational wave (GW) background (due to the mergers of PBH binaries at high-$z$, in the matter dominated era)~\cite{Wang:2016ana}, quasar microlensing~\cite{Mediavilla:2017bok}, lensing of type-Ia supernovae~\cite{Zumalacarregui:2017qqd}, or the orbital dynamics of halo wide binaries~\cite{Monroy-Rodriguez:2014ula}. When approaching the $\sim100\,M_\odot$ scale, radio and X-ray observations of the Milky Way~\cite{Manshanden:2018tze}, the half-light radius of dwarf galaxies~\cite{Brandt:2016aco,Li:2016utv} or the stellar distribution of dwarf galaxies~\cite{Koushiappas:2017chw}  take over as more and more stringent bounds. 
Basically, the CMB constraints surpass all these at masses $M\gtrsim 20-50\,M_\odot$, and remain the dominant constraint until at least $10^{3.5}\,M_\odot$, when they become comparable to (or slightly better than) BBN ones~\cite{Nakama:2014vla,Jeong:2014gna,Inomata:2016uip}, before being definitely surprassed by CMB spectral distortions (see~\cite{Kohri:2014lza,Carr:2018rid} and Refs. therein) at~$M\gtrsim 10^{4.5}\,M_\odot$. Needless to say, since different constraints are derived in different systems and are affected by different systematics, the existence of multiple arguments excluding some parameter-space strengthens their credibility and robustness. In particular, over all the stellar mass range, PBH as the totality of DM are excluded by at least two arguments, often more. It is worth noting that even in the most conservative case that we consider, the CMB now provides an independent argument excluding PBH of $M\gtrsim 15\,M_\odot$ as the totality of DM; around $30\,M_\odot$, no more than $f_{\rm PBH}\sim {\cal O}(0.1)$ is allowed. On the other hand, once accounting for uncertainties, the CMB is not capable of disproving a primordial origin of the bulk of LIGO/Virgo merger events, estimated according to~\cite{Sasaki:2016jop,Ali-Haimoud:2017rtz}. But it is interesting to see how the disk accretion scenario is in strong tension with this interpretation, thus providing a phenomenological motivation for reducing accretion-related uncertainties.

\section{Implications for SMBH}\label{SMBH}
In this section, we discuss the implications, if any, of the CMB bounds previously derived for the still mysterious genesis of SMBH.
\subsection{The ``SMBH problem''}\label{SMBHprob}
Supermassive black holes (SMBH)---loosely defined as BH whose mass exceeds $10^5\,M_\odot$---are believed to sit at the center of almost all galaxies, and their integrated accretion disk emission almost saturates the cosmological X-ray background, see for instance the review~\cite{Merloni:2015dda}. Additionally, SMBH---including a few very massive ones with $M\gtrsim 10^9\,M_\odot$---have been observed at high redshift $z\gtrsim 6$, a fact which seriously constraints their formation mechanism~\cite{2010A&ARv..18..279V}. 

One may wonder if these SMBH may have developed from lighter black holes via accretion phenomena, a process that is known to be at play  at $0\leq z\lesssim 6$. Since SMBH with $M\gtrsim 10^9\,M_\odot$ were already in place by the time the universe was 1 billion years old, they raise some concerns. The standard argument goes as follows:

Based on Eqs.~(\ref{Ledd},\ref{salpetertau},\ref{defepsilon}),  
if $\epsilon$ can be estimated, $L_E$ can be linked to {\it what is considered an optimistic benchmark accretion value}, since we expect the maximal~\footnote{This is probably unrealistic, since it excludes any sizable outflows.} {\it mass} accretion rate to be the complement of mass-accretion rate converted into radiation at Eddington limit, i.e. 
\begin{equation}
 \dot{M}\lesssim \dot{M}_E\equiv\frac{1-\epsilon}{\epsilon}\frac{L_E}{c^{2}}\propto \frac{1-\epsilon}{\epsilon}M\,.\label{accretionnooutfl}
 \end{equation}
Note that the Eddington mass accretion rate $\dot{M}_E$ depends on the unknown quantity $\epsilon$, and is mathematically unbounded from above, when $\epsilon\to 0$, while it can be arbitrarily small, when $\epsilon\to 1$. However, the largest known values of $\epsilon$ that can be attained are $\epsilon\simeq 0.42$ for a maximally rotating Kerr BH, so that the following inequality 
\begin{equation}
\dot{M}_E\gtrsim \dot M_{E,{\rm min}}\simeq \frac{L_E}{c^{2}}=\frac{M}{\tau_E} \,,\label{meddmin}
 \end{equation}
seems to hold for all known systems. 

In the literature, the mass accretion rate of a BH of mass $M_i$ at formation time $t_i$ is thus limited as
\begin{equation}
M(t)\lesssim M_i\times \exp\left(\frac{1-\epsilon}{\epsilon}\frac{t-t_i}{\tau_E}\right) \,,
\end{equation}
where $\tau_E$ was given in Eq.~(\ref{salpetertau}).
For a typical benchmark value $\epsilon\approx 0.1$, this implies a $e$-fold time for the BH of the order of 0.04 Gyr, or a maximum estimated growth by accretion of $17\,e-$folds between the epoch of first stars dying at $z\simeq 15$ and observations at $z\simeq 6$. Thus, there is barely the time for a stellar BH of mass ${\cal O}(100)\,M_\odot$
to grow to the size of the heaviest SMBH at $z\gtrsim 6$.  

The above argument hides the loophole that, if the efficiency of conversion of accretion into luminosity drops, the actual mass growth may be significantly larger. Radiatively inefficient accretion at very high inflow rates has been discussed in the past as a way to evade the above argument, and suggested by a number of simulations and theoretical arguments, see for instance Ref.~\cite{Novak:2013dya,Volonteri:2014lja}. In general, accretion in this regime may become non-stationary, see for instance the considerations in Ref.~\cite{Ricotti:2007jk}. There are also some mechanisms alternative to the accretion mechanism onto stellar-mass black holes to overcome the difficulty in forming SMBH, such as invoking runaway mergers in dense clusters~\cite{1990ApJ...356..483Q,PortegiesZwart:2004ggg}, or direct collapse of BH from a gas cloud~\cite{1994ApJ...432...52L}, see Ref.~\cite{Begelman:2006db} for a review. 

Nonetheless, the difficulty of achieving the required conditions has led several authors (see for instance  Ref.~\cite{Duechting:2004dk}) to speculate that SMBHs or rather of their seeds may have a primordial origin, possibly linked with the origin of galactic structures, an old idea recently reviewed in Ref.~\cite{Carr:2018rid}. 

\subsection{Primordial SMBH?}\label{primSMBH}

Similarly to the Milky Way halo mass fraction in the SGR A* black hole, SMBHs currently account for about $10^{-5}$ of the DM mass density in the universe (e.g. Ref.~\cite{Yu:2002sq}, see also the gray band in Fig. 5 of Ref.~\cite{Shankar:2007zg}). However, as we just reviewed, it is known that SMBH undergo significant growth with time. In fact,  at $z\simeq 6$ the overall mass density into SMBH above $10^6\, M_\odot$  was only about a factor $10^{-3.5}$  of the current value, such that the difference between these figures must be accounted for via mergers, accretion and newly formed objects.  
A more quantitative description of the high-redshift SMBH mass function can be 
given in terms of  the so-called Schechter function,
\begin{equation}
\frac{{\rm d} n_{\rm BH}}{{\rm d} \log_{10} m}=m\ln 10\frac{{\rm d} n_{\rm BH}}{{\rm d}  m}=\kappa\, m^\alpha e^{-m}\,,\label{paramet}
\end{equation}
with inferred values at $z=6$ of $\kappa=1.23\times 10^{-8}{\rm Mpc}^{-3}$, $\alpha=-1.03$ and $m\equiv M/M_*$, with $M_*=2.24\times 10^{9}\,M_\odot$ 
(see Ref.~\cite{2010AJ....140..546W} or equivalently Fig. 2 in Ref.~\cite{2011MNRAS.417.2085V}). 
This is consistent with the inferred co-moving density  $>1.1\times 10^{-9}\,$Mpc$^{-3}$ above  $10^9\,M_\odot$ 
between $z=6.44$ and $z=7.44$ reported in Ref.~\cite{2010AJ....140..546W}. If translated  in terms of the DM fraction, Eq.~(\ref{paramet}) yields about 96$M_\odot\,{\rm Mpc}^{-3}$ above $10^6\, M_\odot$, equivalent to a fraction of the DM abundance in SMBH above $10^6\, M_\odot$ of $f_{\rm PBH}\simeq 2.9\times 10^{-9}$. 
Thus, even under the extreme case of eq.~(\ref{LEconstraint}), {\it the CMB angular power spectra do not exclude a primordial origin hypothesis for the SMBHs.}

Are there counter-arguments to this? An apparent theoretical difficulty is that one expects a direct formation of SMBHs to happen after the weak reaction freeze-out, since the horizon mass scales roughly as $M_H\simeq 10^5(t/{\rm s})\,M_\odot$. However, having a very tiny fraction of matter in the form of Primordial SMBHs at BBN times, or even somewhat after BBN, is not obviously excluded, and only limited by theoretical creativity.
A more serious concern is that, if SMBHs form from (quasi)Gaussian fluctuations, the mass  $6\times10^{4}\,M_{\odot}\lesssim M\lesssim 5\times10^{13}\,M_{\odot}$ is subject to tight constraints coming from CMB spectral distortions~\cite{Kohri:2014lza}. 
No cosmologically relevant abundance is allowed in this range {\it unless} the PBH form out of highly non-Gaussian tail fluctuations~\cite{Nakama:2017xvq,Nakama:2016kfq,Garcia-Bellido:2017aan}.

In summary, this discussion leaves  two possible (primordial) scenarios:
\begin{itemize}
\item[1] {\it Primordial SMBH hypothesis}: SMBHs with a mass function similar to the inferred one, eq.~(\ref{paramet}), are directly of primordial origin.
This requires PBHs to form under rather peculiar highly non-Gaussian conditions in order to fulfill CMB spectral constraints. Also, the bulk of the SMBH population is required to undergo negligible mass growth in the period before reionization (dark ages) not to overshoot the inferred mass function, a condition which appears rather challenging to fulfill and puzzling if compared to the $10^{3.5}$ growth observationally deduced between $z\simeq 6$ and today. 
\item[2]   {\it Primordial SMBH seed hypothesis}: If PBHs form with masses $M\lesssim 10^4\,M_\odot$ and $f_{\rm PBH}\lesssim 10^{-9}$, they are consistent with all present bounds, also for initial Gaussian conditions. The presence of SMBHs with masses $M\gtrsim 10^9\,M_\odot$ at redshift $z\gtrsim 6$ requires then a mass growth by a factor of at least $\simeq 10^5$ or 11.5 $e$-folds, which however 
appears ``easily'' achieved within the naive theory sketched above. A sufficient growth can be attained for $M_{\rm PBH}\simeq 200\,M_\odot$ if accreting at Eddington limit and with $\epsilon=0.1$. A similar mass growth process is needed anyway also in astrophysical scenarios, which are usually further constrained by a later formation epoch (typically $z\lesssim 15$) and, at least for popIII (as opposed to direct collapse) scenarios, by a seed mass not usually exceeding $10^2\, M_\odot$.
\end{itemize}

In the second scenario, note that although the PBH get dressed with a DM halo which is one to two orders of magnitude its mass, the DM halo does not necessarily contribute to the ``inferred'' SMBH mass, since the latter is typically deduced from the properties of the inner accretion disk emission (X-ray or radio data). The DM halo rather creates favorable conditions to boost the baryonic accretion.  Finally, since PBH of the relevant masses can be formed before the earliest cosmological timescales probed (weak reaction freeze-out), it is conceivable that theoretical models are more easily constructed in this scenario. 

\section{Conclusions}\label{concl}
Stellar mass or heavier primordial black holes may have interesting cosmological and astrophysical consequences, even if they only constitute a small fraction of the overall amount of dark matter. However, in such a situation their interplay with the remaining fraction of DM may have peculiar consequences, as it has been noted in several instances (see e.g. Refs~\cite{Lacki:2010zf,Adamek:2019gns,Bertone:2019vsk,Hertzberg:2020hsz} for implications for DM models). Here we have revisited the impact that the growth of DM halos around PBH has on CMB anisotropy constraints. We have elucidated the effects of these DM halos on the baryonic accretion thanks to both simple semi-analytical models and dedicated numerical simulations, and derived state-of-the-art cosmological bounds. These limits are the leading ones in the window between a few tens solar mass (where a number of astrophysical bounds exist, typically at the $f_{\rm PBH}\sim 10^{-3}\to 10^{-1}\,$ level) and the tight cosmological bounds from CMB {\it spectral} distortions at masses $M\gtrsim 10^{4}\,M_\odot$ (for a summary, see Fig.~\ref{fig:summary}). The CMB anisotropy bounds reach very deep down in the $f_{\rm PBH}$ range, becoming as stringent as $\lesssim 10^{-8}$ at the highest masses of applicability, both under the hypotheses of spherical or disk accretion. As the largest uncertainty comes from the accretion model, these bounds could be further refined via dedicated hydro-dynamical simulations in a cosmological setting, which would be particularly useful to explore the role of non-stationarity in SMBH accretion and extend the bounds beyond $10^4\,M_\odot$, but also to assess if the CMB bounds exclude a PBH origin of the LIGO/Virgo merger events. In particular,  analytical calculations predict supersonic motion of the baryons at cosmological scales~\cite{Tseliakhovich:2010bj}; starting from those initial conditions, it would be interesting to understand the dynamics of the baryonic gas actually accreted onto the PBH and their DM halos, subject to shocks and dissipative effects.

Despite still existing uncertainties, we argued that CMB bounds do not prevent a primordial origin for the very heavy supermassive black holes observed already at $z>6$. In particular, we find that a scenario where PBHs with $M\gtrsim 10^3\,M_\odot$ act as the seeds of the SMBH easily fulfills all the known constraints.
A prediction of such a scenario is that rather massive black holes are already around and accreting at $z\sim 30$. Qualitatively, we can thus expect interesting implications for the dark ages, such as non-standard 21 cm and reionization epoch (for first explorations, see e.g.~\cite{Hektor:2018qqw,Mena:2019nhm}), which will surpass next generations CMB observations in constraining e.m. energy injection in the dark ages and therefore definitely deserve additional dedicated studies.

\begin{acknowledgments}
We acknowledge valuable discussions with Yacine Ali-Ha\"{i}moud.
We thank Nagisa Hiroshima for collaboration on an early stage of the project and Nagisa Hiroshima and Guillermo Ballesteros for comments on the manuscript. 
This work is partially supported by the project {\it Multimessenger avenues in gravitational waves} under the program ``Initiatives  de  Recherche Strat\'egique'' - IDEX Univ. Grenoble-Alpes (PDS); by JSPS KAKENHI Grants No.~JP17H01131 (K.K.), MEXT Grant-in-Aid for
  Scientific Research on Innovative Areas JP15H05889, JP18H04594,
JP19H05114 (K.K.), and by WPI, MEXT, Japan  (K.K.), the Toshiko Yuasa France-Japan Particle Physics Laboratory ``TYL-FJPPL'' (VP); by the National Science Foundation under Grant No.~1820861 (DI); by the NYU IT High Performance Computing resources, services, and staff expertise (DI).
\end{acknowledgments}
\bibliography{biblio}

\appendix

\section{Size of the region of influence of a PBH}
\label{ionsizecheck}When computing the constraints on PBH accretion from the CMB, it is assumed that the {\em mean} ionized fraction is affected by the energetic radiation emitted by the PBH. We wish to check whether this assumption is valid, and that one should not consider the influence of PBHs as a local {\em perturbation} to the homogeneous ionization fraction.
It was checked in Ref.~\cite{Ali-Haimoud:2016mbv} that photon can escape the very dense environment close to the PBHs, we therefore focus on the far-away region, where densities are cosmological. As an estimate of the typical size of the region influenced by PBHs, one can calculate the mean free path of emitted keV photons, which deposit their energy mostly via Thomson scattering with non-relativistic electrons, i.e.,
\begin{equation}
    \lambda_T\equiv (n_e \sigma_T)^{-1} \simeq 2\times10^4x_e^{-1}\bigg(\frac{1000}{1+z}\bigg)^3~{\rm pc}\,,
\end{equation}
This is to be compared with the typical distance between PBHs 
\begin{equation}
    \bar{r}=\bigg(\frac{3 M}{4\pi\rho_{\rm PBH}}\bigg)^{1/3}\simeq 2\times10^{-1}\bigg(\frac{M}{f_{\rm PBH}M_{\odot}}\bigg)^{1/3}\frac{1000}{1+z}~{\rm pc}\,.
\end{equation}
Hence, one finds that PBHs can influence (and ionize) all of the region separating them from another PBH as long as
\begin{equation}
    f_{\rm PBH}>10^{-15}x_e^3\frac{M_\odot}{M}\,,
\end{equation}
which is always satisfied given the range of PBH masses and fractions considered in this work.

Additionally, one can calculate the number of PBHs per patch of the CMB sky as seen by {\em Planck} at the smallest multipoles, in order to make sure that these patches average out the contribution of many PBHs to the ionization fraction. In the flat sky approximation, valid at these small scales (large multipole $\ell$), the typical comoving size of the patch
$\lambda\equiv d_A(z_{\rm dec})/\ell\sim10^{10}~{\rm pc}/\ell$ contains
\begin{equation}
    N_{\rm PBH}\sim\frac{a \lambda}{\bar{r}} \simeq 5\times 10^7\ell^{-1}\bigg(\frac{f_{\rm PBH}M_\odot}{M}\bigg)^{1/3}\,.
\end{equation}
At the highest multipole seen by {\em Planck}, $\ell \sim 2000$, there are therefore more than one PBH per patch up to masses $M\sim10^4M_\odot$ for a fraction $f_{\rm PBH}=10^{-9}$.

\section{Dynamical friction of a massive PBH}
\label{friction}
Here we want to assess the cosmological relevance of the dynamical friction that a PBH experiences in the cosmological baryonic gas of density $\rho_b$ moving relatively to it at supersonic speed $v\simeq v_L$ (the drag being suppressed at sub-sonic velocities~\cite{Ostriker:1998fa}). A body of mass $M$ experiences the energy loss rate per unit distance with the same form as for a collisionless medium~\cite{Chandrasekhar:1943ys}
\begin{equation}
-\frac{{\rm d}E}{{\rm d}x}=4\pi\rho_b\frac{(GM)^2}{v^2}\ln \Lambda\,,
\end{equation}
where $\ln \Lambda\simeq {\cal O}(10)$ is the Coulomb logarithm, depending on the ratio of the largest and smallest linear scales involved in the process. 
The energy loss timescale is thus
\begin{equation}
\tau_{\rm loss}=\frac{M v^2/2}{-v\,{\rm d}E/{\rm d}x}=\frac{ v^3}{3G\, M\ln \Lambda}\frac{3}{8\pi G\rho_b}\,,\label{lloss}
\end{equation}
In order to assess its cosmological relevance, let us compute the ratio of the above scale with the Hubble time $\sim H^{-1}$. Plugging $v\simeq v_L(z)$ from Eq.~(\ref{csvL}) and using the first Friedmann equation in Eq.~(\ref{lloss}), we obtain 
\begin{equation}
\tau_{\rm loss}(z)H(z)\simeq 1.8\times 10^{4}\frac{M_\odot}{M}\left(\frac{1+z}{100}\right)^{3/2}\frac{10}{\ln \Lambda}\,.
\end{equation}
This indicates that for stellar-mass PBH this effect is sub-leading at the cosmological epoch of interest, but that at the highest PBH masses of interest  ($M\gtrsim 10^4\,M_\odot$) the baryon-PBH motion may eventually settle to sub-sonic during the dark ages. Hence, for studying intermediate mass PBH accretion in pristine halos, as well as their impact on dark ages observables, this effect should be properly accounted for.  

\end{document}